# Emergence of high-temperature superconducting phase in the pressurized La$_3$Ni$_2$O$_7$ crystals

J. Hou[1,2#], P. T. Yang[1,2#], Z. Y. Liu[1,2#], J. Y. Li[3#], P. F. Shan[1,2], L. Ma[1,4,5], G. Wang[1,2], N. N. Wang[1,2], H. Z. Guo[4,5], J. P. Sun[1,2], Y. Uwatoko[6], M. Wang[3*], G.-M. Zhang[7*], B. S. Wang[1,2*], and J.-G. Cheng[1,2*]

[1]*Beijing National Laboratory for Condensed Matter Physics and Institute of Physics, Chinese Academy of Sciences, Beijing 100190, China*

[2]*School of Physical Sciences, University of Chinese Academy of Sciences, Beijing 100190, China*

[3]*Center for Neutron Science and Technology, Guangdong Provincial Key Laboratory of Magnetoelectric Physics and Devices, School of Physics, Sun Yat-Sen University, Guangzhou, Guangdong 510275, China*

[4]*Key Laboratory of Materials Physics, Ministry of Education, School of Physics and Microelectronics, Zhengzhou University, Zhengzhou 450052, China*

[5]*Institute of Quantum Materials and Physics, Henan Academy of Sciences, Zhengzhou 450046, China*

[6]*Institute for Solid State Physics, University of Tokyo, Kashiwa, Chiba 277-8581, Japan*

[7]*State Key Laboratory for Low dimensional Quantum Physics, Department of Physics, Tsinghua University, Beijing 100084, China*

*E-mails: wangmeng5@mail.sysu.edu.cn; gmzhang@mail.tsinghua.edu.cn; bswang@iphy.ac.cn; jgcheng@iphy.ac.cn

## Abstract

The recent report of pressure-induced structure transition and signature of superconductivity with $T_c \approx 80$ K above 14 GPa in the La$_3$Ni$_2$O$_7$ crystals has garnered considerable attention. To further elaborate this discovery, we carried out comprehensive resistance measurements on the La$_3$Ni$_2$O$_7$ crystals grown with the optical-image floating zone furnace under oxygen pressure (15 bar) by using the diamond anvil cell (DAC) and cubic anvil cell (CAC), which employs the solid (KBr) and liquid (Glycerol) pressure transmitting medium, respectively. For the sample #1 measured in DAC, it exhibits a semiconducting-like behavior with large resistance at low pressures and becomes metallic gradually upon compression. At the pressures $P \geq$ 13.7 GPa, we observed the appearance of resistance drop as large as ~50% around 70 K, which evolves into a kink-like anomaly at pressures above 40 GPa and shifts to lower temperatures gradually with increasing magnetic field. These observations are consistent with the recent report mentioned above. On the other hand, the sample #2 measured in CAC retains the metallic behavior in the investigated pressure range up to 15 GPa. The hump-like anomaly in resistance around ~130 K at ambient pressure disappears at $P \geq$ 2 GPa. In the pressure range from 11 GPa to 15 GPa, we observed the gradual development of a shoulder-like anomaly in resistance at low temperatures, which evolves into a pronounced drop of resistance by 98% below 62 K at 15 GPa, reaching a temperature-independent resistance of 20 μΩ below 20 K. Similarly, this resistance anomaly can be shifted to lower temperatures progressively by applying external magnetic fields, resembling a typical superconducting transition. Measurements on sample #3 in CAC reproduce the resistance drop at pressures above 10 GPa and realize the zero resistance below 10 K at 15 GPa even though an unusual semiconducting-like behavior retains in the normal state. Based on these results, we discussed some issues regarding the sample-dependent behaviors and the effects of pressure condition on the high-temperature superconductivity in $La_3Ni_2O_7$.

## Introduction

The first-row (3d) transition-metal oxides (TMOs) with the perovskite and related structures provide a fertile playground for the discovery of novel emergent quantum phenomena due to the intimated interplay of spin, charge, orbital and lattice degrees of freedom [1-3]. The discovery of unconventional high-temperature superconductivity in cuprates represents one of the most celebrated examples [4-6], and it has thus encouraged continuous effort in finding more unconventional superconducting systems among the 3d TMOs. As the nearest neighbor to copper in the periodic table, the nickel oxides (nickelates) have gained considerable attention as the most promising candidate for high-$T_c$ superconductivity ever since the early 1990s [7-9]. However, experimental breakthroughs in this direction are not achieved until recently. In 2019, Li et al successfully synthesized hole-doped infinite layer $Nd_{1-x}Sr_xNiO_2$ thin films through a topotactic reduction reaction of the perovskite phase by using $CaH_2$ and discovered superconductivity with $T_c$ around 9 ~ 15 K [10-12]. Such a discovery has stimulated many theoretical discussions on the similarities and differences between the cuprates and nickelates [13-16]. It was later found that the $T_c$ of $Pr_{0.82}Sr_{0.18}NiO_2$ thin films can be enhanced to over 30 K at 12.1 GPa, which underscores the potential for further raising $T_c$ of the superconducting nickelates [17].

More recently, Sun et al have reported on the signature of superconductivity with $T_c \approx$ 80 K above 14 GPa in the $La_3Ni_2O_7$ crystal [18], which is the n = 2 member of the Ruddlesden-Popper phases $La_{n+1}Ni_nO_{3n+1}$. Although the polycrystalline $La_3Ni_2O_7$ has been well studied previously, experimental study on the single-crystalline samples is rather scarce due to the difficulty in growing large single crystals [19-22]. By using the optical-image floating zone method, Liu et al [23] synthesized large-sized $La_3Ni_2O_7$ single crystals under a moderate oxygen pressure of $p(O_2)$ = 15 bar and revealed a metallic behavior with obvious anomalies in resistivity and magnetic susceptibility at about 110 and 153 K associated with the formation of a density-wave-like transition in this system. When these $La_3Ni_2O_7$ crystals were subjected to compression in a diamond anvil cell (DAC), Sun et al observed a pressure-induced structural transition from the space group *Amam* to *Fmmm* around 14 GPa, which straightens the interlayer Ni-O-Ni bond angle within the bilayers of $NiO_6$ octahedra [18]. In concomitant with the structural transition, they observed the emergence of a pronounced drop in resistance around 80 K. Since this anomaly can be suppressed gradually by external magnetic fields and the ac magnetic susceptibility also shows a corresponding drop at the similar temperature, it has been attributed to the occurrence of superconducting transition with $T_c \approx 80$ K. This discovery is exciting as it represents the second class of 3d TMO superconductors with $T_c$ above the liquid-nitrogen boiling temperature of 77 K.

To reach a decisive conclusion on the occurrence of high-$T_c$ superconductivity in $La_3Ni_2O_7$, however, some important issues have to be clarified: (i) the zero-resistance state has not been achieved yet; (ii) it remains unclear how the metallic sample at ambient pressure is altered to a semiconductor when subjected to compression in DAC; (iii) how the density-wave-like transitions evolve with pressure and the emergent high-$T_c$ superconductivity also waits for explorations. There is no doubt that more comprehensive high-pressure (HP) studies are needed to address these issues. Previous HP studies on the polycrystalline $La_3Ni_2O_7$ samples have revealed that its transport properties are sensitive to the pressure conditions, i.e., hydrostatic versus non-hydrostatic pressures [24]. In this work, therefore, we carried out a comparative study on the transport properties of $La_3Ni_2O_7$ crystals by using two distinct HP techniques with different pressure conditions, i.e., the DAC and the cubic anvil cell (CAC) employing solid (KBr) and liquid (Glycerol) pressure transmitting medium (PTM), respectively. For the measurements in DAC, the results are consistent with the recent report of Sun et al. [18]. Remarkably, the measurements in CAC with an excellent hydrostatic pressure condition reveal distinct evolutions of the metallic transport properties and the gradual development of resistance drop in the pressure range between 11 GPa and 15 GPa, partially addressing the above-mentioned issues (ii) and (iii). Some sample-dependent issues are also revealed in the present studay. Further effort is needed to measure its resistance under hydrostatic pressures above 15 GPa in order to verify the possible high-temperature superconductivity in $La_3Ni_2O_7$.

## Experimental

Small pieces of $La_3Ni_2O_7$ samples used in the present work were extracted from large crystals grown with the optical-image floating zone furnace under an oxygen pressure of $p(O_2) = 15$ bar at Sun Yat-Sen University in China. Details about the crystal growth and physical-property characterizations at ambient pressure have been published elsewhere [23]. In the present study, we found that the electrical transport properties of the studied $La_3Ni_2O_7$ samples have a strong sample-dependent issue, and the samples can be generally categorized into groups, i.e., some display an insulating behavior over the entire temperature range while others exhibit a metallic behavior with a weak hump-like anomaly in resistance around 110-130 K. Such observations of distinct, sample-dependent transport behaviors should be attributed to the different oxygen stoichiometry in the studied samples. According to the previous studies on the polycrystalline $La_3Ni_2O_{7-\delta}$, its transport property can change significantly from metallic to insulating behavior for δ as small as 0.08 [25]. It is thus plausible that the as-grown $La_3Ni_2O_7$ crystals have non-uniform distribution of oxygen concentration. In addition to the chemical inhomogeneity, the oxygen stoichiometry of $La_3Ni_2O_7$ crystals might also be modified during the post-handling process. We found that the contact resistance of some $La_3Ni_2O_7$ crystals increases when the electrical leads were directly attached on the sample's surface by using sliver paste (DuPont 4929N). To prevent the direct contact of silver paste with sample's surface, we deposited four platinum (Pt) pads with the magnetron sputtering machine and then attached the gold wires on these Pt pads with silver paste.

We performed HP resistance measurements on three $La_3Ni_2O_7$ samples by using the DAC (#1) and CAC (#2 and #3), respectively. For the measurements in DAC, the sample #1 (inset of Fig. 1a) was loaded into a pre-indented gasket hole filled with KBr as the solid PTM in between a pair of diamond anvils with 300 μm culet. Four golden leads were manually put on the sample's surface and the electrical contact is maintained by the mechanical compression. A piece of ruby ball placed near the sample in DAC is used as the pressure calibrant and the pressure is determined by monitoring the position of the ruby fluorescence R1 line at room temperature. For the samples #2 and #3, we employed a palm-type CAC for resistance measurements up to 15 GPa by using a standard four-probe method. The sample #2 has a dimension of $0.60 \times 0.21 \times 0.06$ mm$^3$ and the sample #3 has a dimension of $0.52 \times 0.16 \times 0.09$ mm$^3$. As seen in Fig. 2(a), the sample is hung inside the Teflon capsule filled with glycerol as the liquid PTM. As shown in our previous work, the three-axis compression geometry together with the adoption of liquid PTM ensures an excellent hydrostatic pressure conditions up to 15 GPa in CAC [26]. The pressure values inside the CAC were estimated from the pressure-loading force calibration curve pre-determined by measuring the structure phase transitions of Bi, Sn and Pb at room temperature. All low-temperature experiments were carried out in a $^4$He refrigerated cryostat equipped with a 9 T superconducting magnet.

## Results and discussion

**High-pressure resistance measurements on the sample #1 in DAC**

Figure 1(a) shows the temperature-dependent resistance $R(T)$ in a semi-logarithmic scale for sample #1 measured in DAC under various pressures up to 50.1 GPa. Because this sample is very small, we did not measure its $R(T)$ at ambient pressure. At the first applied pressure of 7 GPa in DAC, the $R(T)$ exhibits a weakly insulating behavior upon cooling down with the room-temperature value of about 25 Ω. The metallic behavior at AP is completely altered upon compression in DAC. Similar observations were also reported in literature [18]. When the pressure is increased to 10.4 GPa, the $R(T)$ decreases in the whole temperature range, exhibits a weaker temperature dependence at high temperatures, and then displays a down-turn feature at about 40 K followed a slight upturn at the lower temperatures. Upon further increasing pressure to 13.7 GPa, a metallic behavior is realized in the $R(T)$ at high temperatures, and a clear drop of resistance by 50% appears at about 70 K, which is close to but slightly lower than that of $T_c$ defined as the superconducting transition temperature in Ref. [18].

Like the previous report [18], a finite value of resistance is preserved below the transition and $R(T)$ even goes up again at temperatures below 20 K. Thus, we define the transition temperature as $T_u$ from the intercept of two straight lines across the resistance anomaly as illustrated in Fig. 1(b). No thermal hysteresis is observed around this anomaly as shown in Fig. 1(b). Similar $R(T)$ behaviors are observed at pressure up to 29.2 GPa, and the transition temperature $T_u$ first increases slightly to 70.7 K at 16.8 GPa and then decreases gradually to 63 K with increasing pressure to 29.2 GPa. When the pressure is increased from 29.2 GPa to 39.8 GPa, the room-temperature resistance decreases suddenly by nearly one order of magnitude from ~4 Ω to 0.4 Ω, and $R(T)$ exhibits a strong temperature dependence in the whole temperature range. Meanwhile, the low-temperature upturn disappears and the drop anomaly at $T_u$ is replaced by a kink-like anomaly in $R(T)$, and $T_u$ is suppressed to about 50 K at 50 GPa. At the highest pressure of 50.1 GPa, we also measured $R(T)$ under different external magnetic fields displayed in Fig. 1(c), which shows a gradual suppression of the resistance below $T_u$, in reminiscent of the observations in Ref. [18]. Overall, the above results are consistent with those reported results [18], showing a pronounced drop or kink-like anomaly in resistance at $T_u \approx 70$ K under pressures above ~13 GPa. Due to the absence of zero resistance for the above data in DAC, however, it was ascribed it to signature of superconducting transition at this stage.

**High-pressure resistance measurements on the sample #2 in CAC**

At AP, the crystal $La_3Ni_2O_7$ shows a metallic behavior with a density-wave-like transition at $T_s \approx 110\text{-}150$ K, which is manifested as obvious anomalies in $R(T)$. For the polycrystalline samples, the temperature-dependent $R(T)$ changes from metallic above $T_s$ to a semiconducting-like behavior below $T_s$ [27]. In contrast, the measured $R(T)$ of $La_3Ni_2O_7$ single crystals presents a weaker hump-like anomaly at $T_s$ and it remains metallic below $T_s$ down to the lowest temperature. Previous HP studies on the transport properties of $La_3Ni_2O_7$ polycrystalline samples have shown that both resistance and $T_s$ decrease monotonically with pressure below 2 GPa with a slope of $dT_s/dP$ = 10.7-12.5 K/GPa, which would predict a complete suppression of $T_s$ at ~10 GPa [24]. For the above HP resistance measurements in DAC, unfortunately, the metallic behavior has been altered by the nonhydrostatic pressure conditions and it thus prevents us from observing intrinsic evolutions of the transport properties under HP. To overcome this problem, we resorted to the CAC technique, which employs a three-axis compression geometry and adopts liquid PTM so as to maintain an excellent hydrostatic pressure conditions up to 15 GPa.

Figure 2(a) shows the obtained $R(T)$ data of sample #2 under various pressures up to 15 GPa in CAC. At AP, $R(T)$ displays a metallic behavior in the whole temperature range with a weak hump-like anomaly at $T_s$ = 136 K, which can be determined from the peak position of $dR/dT$. In comparison with the previous report showing $RRR$ = 2.5, the present sample exhibits a larger $RRR$ = 15, signaling a better quality of the sample. The room-temperature resistivity value is estimated to be ~18 mΩ cm. As seen in Fig. 2(a), the metallic behavior of $R(T)$ is retained upon compression in CAC, strikingly different from the above results in DAC, and the magnitude of resistance decreases monotonically with increasing pressure, especially in the high-temperature region. At the first applied pressure of 2 GPa, the hump-like anomaly in $R(T)$ cannot be discerned any more, suggesting the complete suppression of the density-wave-like transition by pressure. To track the evolution of $T_s(P)$ in the low-pressure range, HP resistance measurements by using a piston-cylinder cell are in progress.

For the pressures between 2 GPa and 9 GPa, the $R(T)$ curves show a plain metallic behavior without any discernable anomaly down to 1.5 K. Upon further increasing pressure to 11 GPa, we observed in $R(T)$ a slight upturn below 10 K followed by a faint drop below 5 K, which even shows some field dependences. This feature motived us to measure $R(T)$ in a finer pressure interval from 11 to 15 GPa. As seen in Fig. 2(b), the measured $R(T)$ at 12 GPa exhibits a weak kink around 25 K and then develops a broad shoulder centered around 10 K, which can be eliminated by external magnetic fields, (see Fig. S1). These features become more pronounced and moves to higher temperatures with increasing pressure, and the shoulder feature also fades away gradually. Finally, the resistance $R(T)$ at 15 GPa is characterized by a quite sharp drop at $T_0 \approx$ 62 K followed by a long tail between ~48 K and 20 K and then a nearly temperature-independent resistance of 20 μΩ below 20 K. This value is about 2% of the resistance at $T_0$, i.e., the resistance drops by 98% from $T_0$ to the lowest temperature of 1.5 K. Such a large drop of resistance with a nearly temperature-invariant $R(T)$ over a wide temperature range is unlikely caused by magnetic or electronic phase transitions but is more likely associated with the occurrence of superconducting transition. The presence of a small residual resistance together with the observation of a relatively wide transition indicates that the volume ratio of the possible superconducting phase is low at this pressure. Indeed, measurements of ac magnetic susceptibility up to 15 GPa in CAC fails to detect any superconducting shielding effect, in accordance with the resistance results. Upon decompression from 15 GPa, the recovered $La_3Ni_2O_7$ sample keeps intact and its $R(T)$ is measured again at AP. As shown in Fig. 2(a), the obtained $R(T)$ is almost identical to the initial curve before compression. This result confirms that the pressure-induced changes are reversible at least up to 15 GPa in CAC.

To check whether the observed resistance drop is truly associated with a superconducting transition, we measured $R(T)$ at 15 GPa under different magnetic fields. Figure 2(c) shows the normalized $R(T)$ below 65 K. As can be seen, the transition shifts to lower temperatures gradually with increasing magnetic field, similar to a superconducting transition. Thus, we tentatively ascribed to the observed resistance drop to the onset of superconducting transition, i.e., $T_0 = T_c^{onset}$. Here we define $T_c^{90\%}$ and $T_c^{50\%}$ at each field according to the criteria of 90% and 50% of the corresponding normal-state resistance at $T_c^{onset}$, and then plotted the temperature dependence of $\mu_0 H_{c2}(T_c)$ in Fig. 2(d). Least-square fitting to these data by using the empirical Ginzburg-Landau (GL) equation yields the zero-temperature upper critical field $\mu_0 H_{c2}(0)$ = 92 T and 43 T for $T_c^{90\%}$ and $T_c^{50\%}$, respectively.

**High-pressure resistance measurements on the sample #3 in CAC**

To check the reproducibility of the observed resistance drop in the sample #2, we loaded another sample (#3) in CAC for resistance measurements at high pressures. As seen in Fig. 3(a), this sample shows a metallic behavior over the entire temperature range at AP, with an obvious anomaly at about $T_s \approx 110$ K in $R(T)$. The metallic behavior can be retained at 2 GPa, but changes to a semiconducting behavior at 10 GPa, which is different from the results of the sample #2 (Fig. 2(a)). As can be seen, the measured $R(T)$ at 10 GPa first increases upon cooling down with an obvious slope change around 150 K, and then drops suddenly below ~20 K. At this pressure the resistance drops by 90% at 1.5 K and the applied magnetic fields can suppress this drop, signaling the occurrence of a superconducting transition. With increasing pressures from 10 GPa to 12 GPa,14 GPa and 15 GPa, the resistance in the normal state retains the similar feature but the magnitude decreases progressively, while the drop in resistance moves quickly to higher temperatures, with the onset temperature increasing to ~ 42 K, 60 K and 64 K at 12 GPa, 14 GPa and 15 GPa, respectively. Interestingly, the resistance $R(T)$ at 14 GPa and 15 GPa can finally reach zero below about 5 K and 10 K (see the lower panel of Fig. 3(b)), confirming the occurrence of superconductivity. However, the observation of such a broad transition indicates that the superconducting volume fraction is still low at this pressure. As shown in Fig. 3(c), the field dependence of $R(T)$ is also consistent with the superconducting transition.

**Temperature-pressure phase diagram**

The characteristic temperatures $T_s$, $T_u$, and $T_0$ defined in the above DAC and CAC resistance measurements as a function of pressure are summarized in Fig. 4. The $T_c$ values determined from the measurements in DAC in Ref. [18] are also included for comparison. As can be seen, the $T_u(P)$ determined in the present work matches well to the value of $T_c(P)$ of the previous work, and the $T_0(P)$ connects rather smoothly to the $T_c(P)$, implying a common origin for these anomalies in resistance. Importantly, our results together with the previous study reveal a dome-shaped superconducting phase in the pressurized $La_3Ni_2O_7$ crystals. Whether it is associated with the proposed structural phase transition or adjacent to other electronic phases deserves further experimental investigations.

Through the above detailed resistance measurements in CAC, we have revealed the emergence and gradual development of a possible superconducting phase above 10 - 11 GPa, and its volume seems to increase gradually with pressure but remains in a filamentary nature up to 15 GPa. This might be related to the nature of a first-order structure transition around 15 GPa from the low-pressure phase (space group *Amam*) to the HP phase (space group *Fmmm*) as reported in the recent experiment [18]. Usually, the coexistent phases appear in the first-order transition, and span over a finite region in the phase diagram. More studies are certainly needed to check whether bulk superconducting state can be eventually achieved at higher pressure in $La_3Ni_2O_7$. In this regard, a two-stage multianvil press that can reach hydrostatic pressures up to 20 GPa is valuable for this purpose. Nonetheless, the present study provides the detailed information about the emergence and evolution of a superconducting phase in the $La_3Ni_2O_7$ crystal in the critical pressure regime.

**The sample-dependent issues**

In the present study, we have provided more evidence in support of the emergence of high-$T_c$ superconducting phase above ~10 GPa in the $La_3Ni_2O_7$ crystals grown under a moderate oxygen pressure (15 bar). As mentioned above, however, our results also suggest strong sample-dependent issues, presumably associated with the variations of oxygen content in the samples, in addition to the sensitivity to the pressure environments. For the samples #2 and #3 measured in CAC, both show a metallic behavior with a weak hump-like anomaly in $R(T)$, but their $T_s$ = 136 K and 103 K are slightly different. In addition, the sample #3 possess much reduced *RRR* = 2.5 and its room-temperature resistance is about one order higher than that of sample #2 as shown in Fig. 2(a). These differences might be responsible for the observed distinct responses of resistance to pressures at high temperatures. Under high pressures, the sample #2 retains a metallic behavior up to 15 GPa, while the sample #3 changes into a semiconducting behavior. Since the pressure conditions are the same, it remains unclear why these two samples display distinct response to the applied pressure. More detailed studies on the samples with well-controlled quality are needed to finally resolve these issues.

## Conclusion

In summary, we have employed a palm-type CAC apparatus to measure the HP resistance of $La_3Ni_2O_7$ crystals up to 15 GPa. Under hydrostatic pressure conditions, our results confirm the emergence of superconducting phase in $R(T)$ at pressure above 10-11 GPa, and the onset temperature can reach about 62 K at 15 GPa. Together with the present and previous results in DAC, the present study reveals a dome-shaped superconducting phase in the pressurized $La_3Ni_2O_7$ crystals.

## Acknowledgements

This work is supported by the National Key Research and Development Program of China (2018YFA0305700, 2021YFA1400200), the National Natural Science Foundation of China (12025408, 11921004), the Beijing Natural Science Foundation (Z190008), the Strategic Priority Research Program of CAS (XDB33000000), and the Users with Excellence Program of Hefei Science Center CAS (2021HSC-UE008). The high-pressure experiments were performed at the Cubic Anvil Cell station of Synergic Extreme Condition User Facility (SECUF). Work at Sun Yat-Sen University was supported by the National Natural Science Foundation of China (12174454), Guangdong Basic and Applied Basic Research Funds (2021B1515120015), and Guangdong Provincial Key Laboratory of Magnetoelectric Physics and Devices (2022B1212010008).

## Reference

[1]. D. Adler, Mechanisms for metal-non-metal transitions in transition-metal oxides and sulfides, *Rev. Mod. Phys*. **40**, 714 (1968).

[2]. M. Imada, A. Fujimori & Y. Tokura, Metal-insulator transitions, *Rev. Mod. Phys*. **70**, 1039 (1998).

[3]. J.G. Bednorz and K. A. Muller. Possible high $T_c$ superconductivity in the Ba−La−Cu−O system. *Z. Physik B - Condensed Matter* **64**, 189–193 (1986).

[4]. P. W. Anderson. The Resonating Valence Bond State in La2CuO4 and Superconductivity. *Science* **235**,1196-1198 (1987).

[5]. P. A. Lee *et al*. Doping a Mott insulator: Physics of high-temperature superconductivity. *Rev. Mod. Phys*. **78**, 17-85 (2006).

[6]. Keimer B. *et al*. From quantum matter to high-temperature superconductivity in copper oxides. *Nature* **518**, 179–186 (2015).

[7]. M. A. Hayward *et al*. Sodium Hydride as a Powerful Reducing Agent for

Topotactic Oxide Deintercalation: Synthesis and Characterization of the Nickel Oxide LaNiO2. *J. Am. Chem. Soc.* **121**, 8843-8854 (1999).

[8]. A. V. Boris *et al*. Dimensionality Control of Electronic Phase Transitions in Nickel-Oxide Superlattices. *Science* **332**, 937-940 (2011).

[9]. A. S. Disa *et al*. Orbital Engineering in Symmetry-Breaking Polar Heterostructures. *Phys. Rev. Lett.* **114**, 026801 (2015).

[10]. D. Li *et al*. Superconductivity in an infinite-layer nickelate. *Nature* **572**, 624-627 (2019).

[11]. M. Osada *et al*. Phase diagram of infinite layer praseodymium nickelate $Pr_{1-x}Sr_xNiO_2$ thin films. *Phys. Rev. Mater.* **4**, 121801(R) (2020).

[12]. G. A. Pan *et al*. Superconductivity in a quintuple-layer square-planar nickelate. *Nat. Mater.* **21**, 160–164 (2022).

[13]. K.-W. Lee & W. E. Pickett. Infinite-layer $LaNiO_2$: $Ni^{1+}$ is not $Cu^{2+}$. *Phys. Rev. B* **70**, 165109 (2004).

[14]. H. Sakakibara *et al*. Model Construction and a Possibility of Cupratelike Pairing in a New $d^9$ Nickelate Superconductor $(Nd,Sr)NiO_2$. *Phys. Rev. Lett.* **125**, 077003 (2020).

[15]. A. S. Botana & M. R. Norman. Similarities and Differences between $LaNiO_2$ and $CaCuO_2$ and Implications for Superconductivity. *Phys. Rev. X* **10**, 011024 (2020).

[16]. G.-M. Zhang, Y.-f. Yang & F.-C. Zhang, Self-doped Mott insulator for parent compounds of nickelate superconductors, *Phys. Rev. B* **101**, 020501(R) (2020).

[17]. N.-N. Wang *et al*. Pressure-induced monotonic enhancement of $T_c$ to over 30 K in superconducting $Pr_{0.82}Sr_{0.18}NiO_2$ thin films. *Nat. Commun.* **13**, 4367 (2022).

[18]. H. Sun, et al. Signatures of superconductivity near 80 K in a nickelate under high pressure. *Nature* https://doi.org/10.1038/s41586-023-06408-7 (2023).

[19]. Z. Zhang, Greenblatt M. & Goodenough J. B. Synthesis, Structure, and Properties of the Layered Perovskite $La_3Ni_2O_{7-\delta}$. *J. Solid State Chem.* **108**, 402-409 (1994).

[20]. K. Yoshiaki *et al*. Transport and Magnetic Properties of $La_3Ni_2O_{7-\delta}$ and $La_4Ni_3O_{10-\delta}$. *J. Phys. Soc. Jpn.* **65**, 3978 (1996).

[21]. T. Satoshi *et al*. Transport, Magnetic and Thermal Properties of $La_3Ni_2O_{7-\delta}$. *J. Phys. Soc. Jpn.* **64**, 1644 (1995).

[22]. M. Greenblatt *et al*. Electronic properties of $La_3Ni_2O_7$ and $Ln_4Ni_3O_{10}$, Ln=La, Pr and Nd. *Synthetic Metals* 85, 1451-1452 (2007).

[23]. Z. Liu *et al*. Evidence for charge and spin density waves in single crystals of $La_3Ni_2O_7$ and $La_3Ni_2O_6$. *Sci. China-Phys. Mech. Astron.* **66**, 217411(2023).

[24]. T. Hosoya et al. Pressure studies on the electrical properties in $R_{2-x}Sr_xNi_{1-y}Cu_yO_{4+\delta}$ (R=La, Nd) and $La_3Ni_2O_{7+\delta}$. *J. Phys.: Conf. Ser.* **121**, 052013 (2008).

[25]. S. Taniguchi et al. Transport, Magnetic and Thermal Properties of $La_3Ni_2O_{7-\delta}$. *J. Phys. Soc. Jpn.* **64**, 1644-1650 (1995).

[26]. J.-G. Cheng *et al*. Cubic anvil cell apparatus for high-pressure and low-temperature physical property measurements. *Chinese Phys. B* **27** 077403 (2018).

[27]. G. Wu *et al*. Magnetic susceptibility, heat capacity, and pressure dependence of the electrical resistivity of $La_3Ni_2O_7$ and $La_4Ni_3O_{10}$. *Phys. Rev. B* **63**, 245120 (2001).

[28]. H. Zheng, et. al., High pO2 floating zone crystal growth of the perovskite nickelate PrNiO3, Crystals 9, 324 (2019).

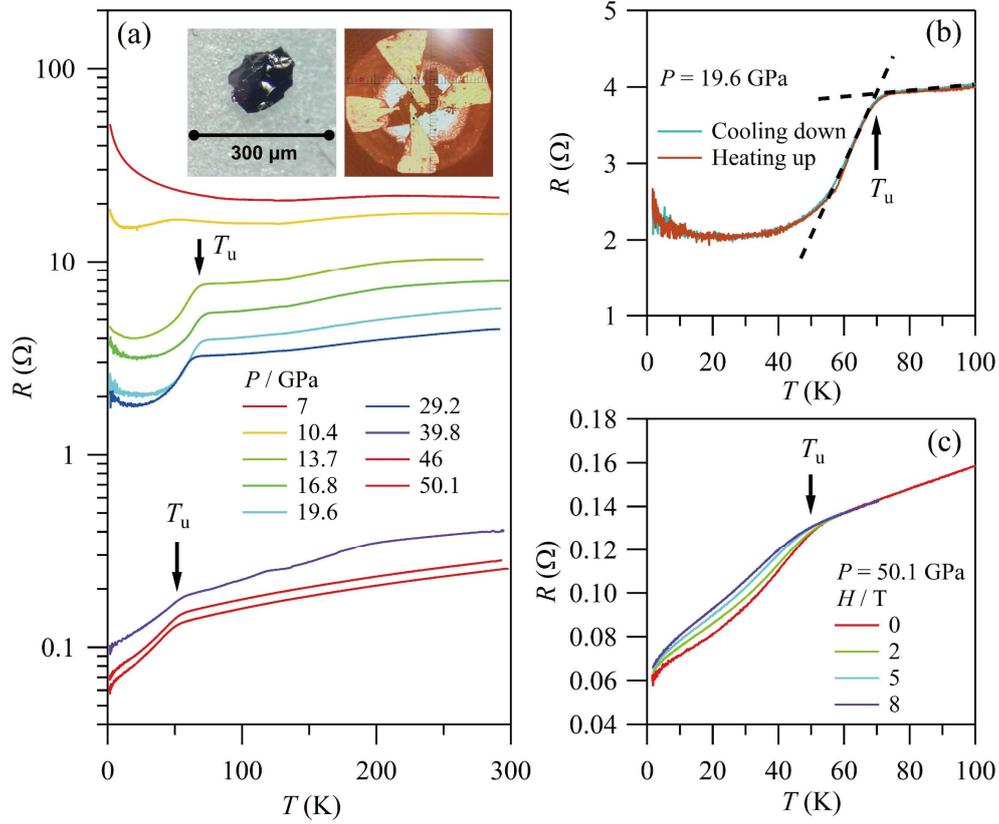

**Figure 1** (a) Temperature-dependent resistance $R(T)$ of $La_3Ni_2O_7$ (#1) crystal under various pressures up to 50.1 GPa measured in a DAC employing KBr as the solid pressure transmitting medium; Inset shows the images of the crystal at AP and in the DAC; (b) An enlarged view of the $R(T)$ at 19.6 GPa below 100 K recorded upon warming-up and cooling-down processes; (c) $R(T)$ curves at 50.1 GPa under different magnetic fields from 0 to 8 T.

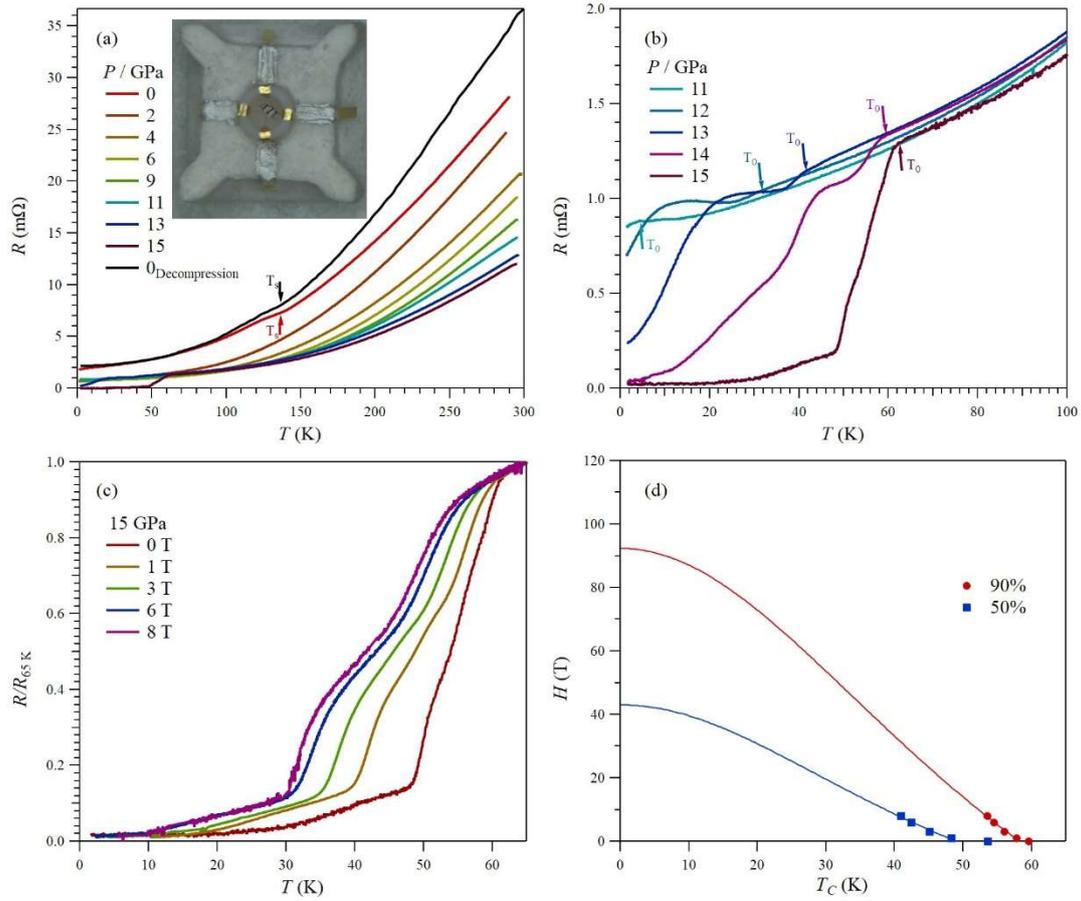

**Figure 2** (a) Temperature-dependent resistivity $R(T)$ of La$_3$Ni$_2$O$_7$ (#2) crystal under various hydrostatic pressures up to 15 GPa measured in a CAC employing glycerol as the liquid pressure transmitting medium; (b) The enlarged view of low-temperature $R(T)$ data at pressures between 11 and 15 GPa, highlighting the gradual development of the resistance drop upon compression; (c) The normalized low-temperature $R(T)/R(65K)$ at 15 GPa under different magnetic fields; (d) The temperature dependences of $\mu_0 H_{c2}(T)$ fitted by the Ginzburg-Landau (GL) equation.

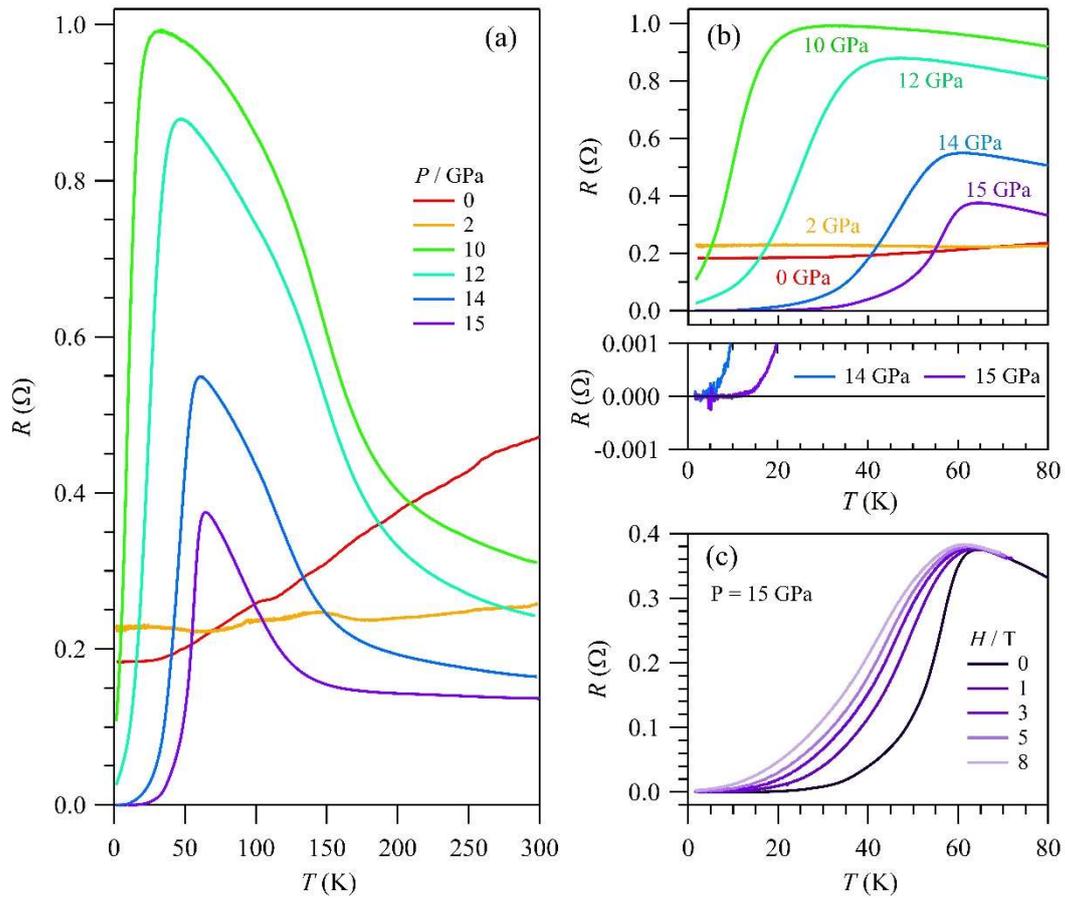

**Figure 3** (a) Temperature-dependent resistivity $R(T)$ of La$_3$Ni$_2$O$_7$ (#3) crystal under various hydrostatic pressures up to 15 GPa measured in a CAC employing glycerol as the liquid pressure transmitting medium; (b) The enlarged view of low-temperature $R(T)$ data; (c) The low-temperature resistivity $R(T)$ at 15 GPa under different magnetic fields.

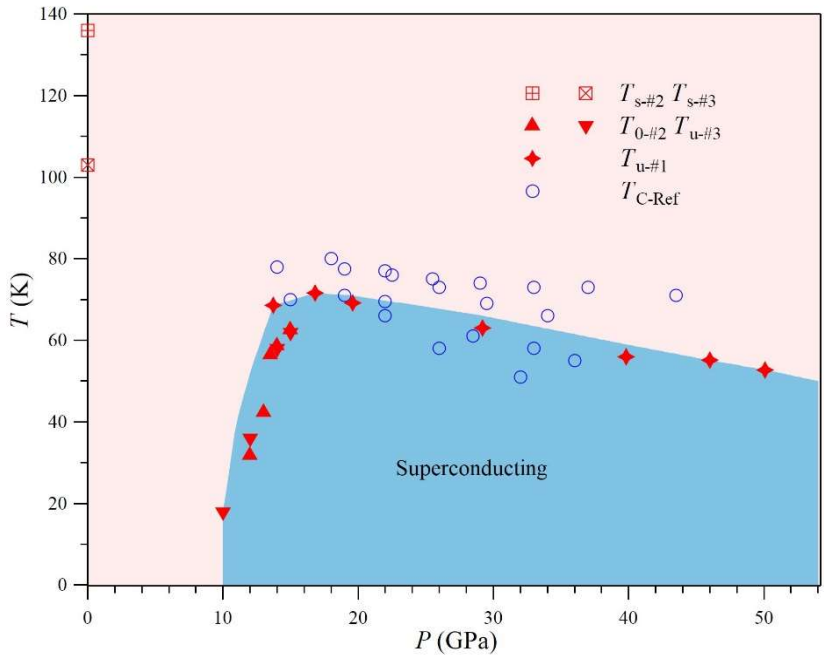

**Figure 4** The *T-P* phase diagram of $La_3Ni_2O_7$ crystal. The red squares represent the density-wave-like transition $T_s$ measured at ambient pressure. The red filled triangles and stars represent the onset superconducting temperatures determined from the present measurements with CAC and DAC, respectively. The blue circles are taken from Ref. [18].

# Supplementary Information

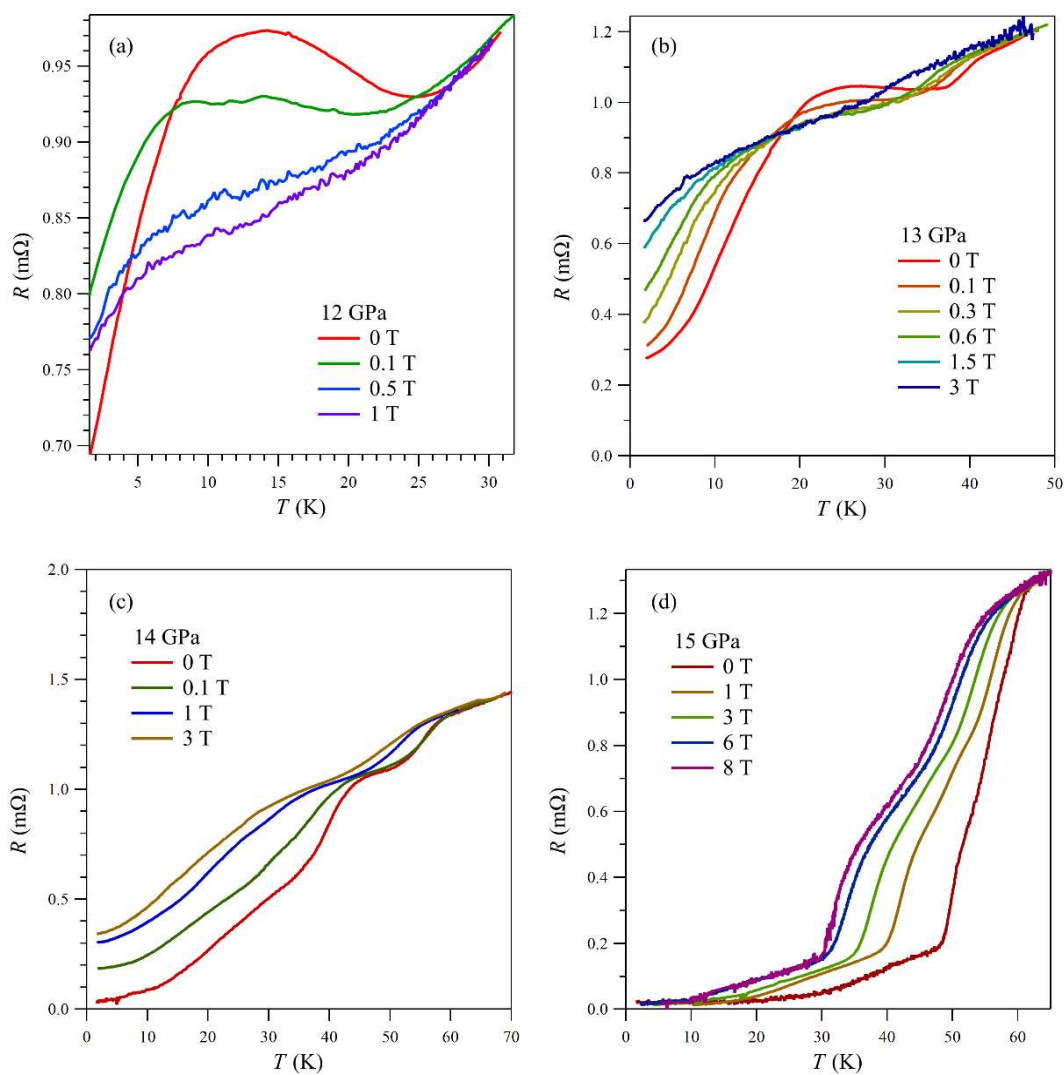

**Figure S1** The low-temperature resistance of $La_3Ni_2O_7$ (#2) crystal under different magnetic fields at: (a) 12 GPa; (b) 13 GPa; (c) 14 GPa; (d) 15 GPa measured in CAC.